\documentclass[12pt,preprint]{aastex}

\shorttitle{Do radio magnetars J1550$-$5418 and J1622$-$4950 have
gigahertz-peaked spectra?}
\shortauthors{Kijak et al.}

\begin{document}


\title{Do radio magnetars J1550$-$5418 and J1622$-$4950 have gigahertz-peaked spectra?}


\author{J. Kijak, L. Tarczewski, W. Lewandowski and G. Melikidze\altaffilmark{1}}
\affil{Kepler Institute of Astronomy, University of Zielona G\'ora, \\
Lubuska 2, 65-265 Zielona G\'ora, Poland}
\email{jkijak@astro.ia.uz.zgora.pl}

\altaffiltext{1}{Abastumani Astrophysical Observatory, Ilia State University, 3-5 Cholokashvili Ave., Tbilisi, 0160,
Georgia}

\begin{abstract}
We study the radio spectra of two magnetars PSR J1550$-$5418 and J1622$-$4950. 
We argue that they are good candidates for the pulsars with gigahertz-peaked 
spectra as their observed flux density decreases at frequencies below 7~GHz.
We suggest this behavior is due to the influence of pulsars' environments on 
the radio waves. Both of the magnetars are associated with supernova remnants, 
thus are surrounded by a hot, ionized gas, which can be responsible for the free-free
absorption of radio-waves. We conclude that the GPS feature of both magnetars and 
typical pulsars are formed by similar processes in the surrounding media, 
rather than by different radio-emission mechanisms. Thus, the radio magnetars 
PSR J1550$-$5418 and J1622$-$4950 can be included in
the class of the gigahertz-peaked spectra pulsars.
\end{abstract}





\section{Introduction}
Most of the radio pulsars are observed to have simple spectra which can be described by a power-law function with an
average spectral index of $-1.8$ \citep{maron00}. Some sources, moreover, show a low-frequency turnover, normally
observed at frequencies below a few hundred megahertz \citep{sieb73}. On the other hand there have been observed few pulsars whose spectra are observed to peak above 1~GHz followed by an otherwise typical high-frequency spectrum
\citep{kijak11a}. At frequencies below 1~GHz the observed flux decreases and the corresponding spectral index becomes
positive. Recently, while analyzing a distribution of spectral indices, \citet{bates13} concluded that the fraction of
GPS pulsars might be less than 10\% of the whole population. In order to describe the pulsars showing these features
\citet{kijak11a} suggested the term gigahertz-peaked spectra (GPS) pulsars. It seems that the appearance of this type of
spectrum is related to the peculiar environments (such as PWNs, HII regions, etc.) of these neutron stars. \citet{kijak11a}
also indicated that the GPS pulsars are relatively young objects and may coincide with the X-ray sources from 3rd EGRET
Catalogue or HESS observations. Thus, they suggested that the GPS occurs not due to the radio emission mechanism but
due to influence of pulsars' environments on the radio waves. Perhaps the most convincing evidence of this statement is a
unique binary system of PSR B1259$-$63 and Be star LS 2883 recently studied by \citet{kijak11b}. They showed that the spectral shape depends on the orbital phase, which definitely proves that the Be star environment is the main reason for PSR B1259$-$63 spectral evolution. The spectrum of B1259$-$63 at
various orbital phases mimics the spectrum of GPS pulsars.
\citet{kijak11b} considered two mechanisms
that might influence the observed radio emission: free-free absorption and cyclotron resonance.

\begin{figure*}
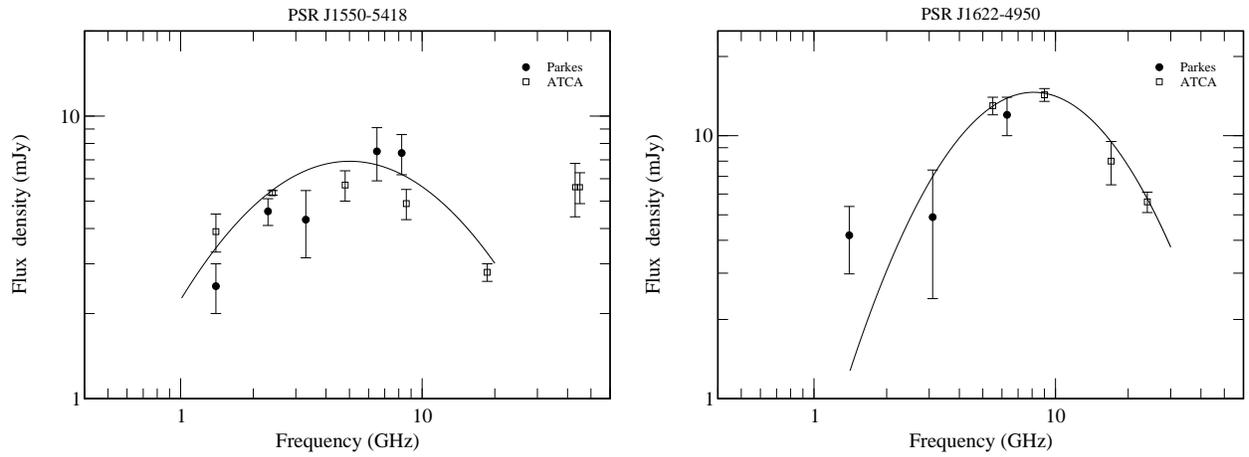

\begin{tabular}{lr}
{\mbox{\includegraphics[width=80mm,angle=0.]{fig1a.eps}}}&
{\mbox{\includegraphics[width=80mm,angle=0.]{fig1b.eps}}}
\end{tabular}
\caption{The average spectra of PSR J1550$-$5418 and J1622$-$4950 obtained from published data: Black dots represent the flux
measurements obtained using Parkes radio telescope and open squares represent flux measurements obtained by the Australia
Telescope Compact Array. Data are taken from \citet{cam07, cam08}, \citet{bates11} for J1550$-$5418 and \citet{lev10, lev12},
\citet{keith11} and \citet{and12} for J1622$-$4950.} \label{fig1}
\end{figure*}

A periodic signal from PSR J1550$-$5418 with the rotation period of 2.069 seconds was first detected on June 8, 2007 by the
Parkes telescope in Australia at 1374 MHz \citep{cam07}. It has a characteristic age of 1.4 kyr and the surface magnetic field
of $2.2~\times~10^{14}$~G. The dispersion measure DM $=830$ cm$^{-3}$pc and the corresponding distance is 9.74 kpc. However,
\cite{del12} using the VLBI measurements estimated a likely distance of $6\pm2$ kpc to this pulsar. PSR J1550$-$5418 is also a
X-ray source 1E 1547.0$-$5408 identified as a magnetar candidate by \cite{gelf07} and later confirmed to be the magnetar by
\cite{cam07}. Its period (2~s) and the spin-down luminosity $\dot E = 1~\times~10^{35}$ ergs~s$^{-1}$ cause that PSR
J1550$-$5418 has the shortest $P$ and highest $\dot E$ among the known magnetars. It also has a peculiar environment \citep[see
details in][]{gelf07}. \citet{cam08} analysed the spectrum of PSR J1550$-$5418 in a wide frequency range (1.4~-~45 GHz) and
suggested that the spectrum could be described as approximately log parabolic with a peak at $\approx 6$ GHz. Later
\citet{bates11} inscribed PSR J1550$-$5418 in the list of pulsars with positive spectral index in a frequency range between 1.4
and 7 GHz. Therefore, J1550$-$5418 seems to be a natural candidate to be classified as the GPS pulsar.

The PSR J1622$-$4950, the radio magnetar with a period of 4.3~s and spin-down luminosity $\dot E = 8.5~\times~10^{33}$
ergs~s$^{-1}$, was discovered by \cite{lev10}. An integrated column density of free electrons DM$=820$ cm$^{-3}$pc, therefore
the corresponding distance to this pulsar is about 9.14~kpc, thus its parameters are similar to those of PSR J1550$-$5418. This
radio magnetar is also a young (4 kyr), highly magnetized neutron star ($>10^{14}$~G) and is referred to as an anomalous X-ray
pulsar. \cite{and12} suggested that the object G333.9+0.0 is a supernova remnant physically associated with PSR J1622$-$4950.
Having analyzed the spectrum between 1.4 and 9 GHz \cite{lev10} claimed that a spectral index might be positive. However,
\cite{keith11} combined the flux density measurements up to 24 GHz and concluded that the spectral index of PSR J1622$-$4950 is
very close to zero.

Here we analyze spectral shapes
of two radio magnetars J1550$-$5418 and J1622$-$4950 and suggest that these magnetars are good candidates for the objects with GPS.

To trace the origin of GPS we need to assume that the radio emission mechanisms for both  magnetars and normal pulsars
share the same feature, namely the coherent curvature radiation \citep{mgp00,gil04,mgm09,rea12}. Consequently the average
spectra of magnetars and normal pulsars should have some similar features too, which might be the power-law spectra with the
turnover at low frequencies. Generally, the radio-emission features of pulsars and magnetars differ significantly, e.g. the
profiles of magnetars are not stable, the position angle varies in a different way, etc. But according to the observations they
have at least one common feature, highly linearly polarized single pulses with the position angle following locally the mean
position angle traverse are observed from both normal pulsars and magnetars \citep{lev12, mgs11, kram2007}. We believe that
this detail is of a high importance. It points to the curvature radiation as the main mechanism for the radio-emission
\citep{mgp00,gil04,mgm09}. The difference can be explained by the relaxation of the twisted magnetosphere, which in general,
does not have a simple dipolar geometry responsible for both the stable shape of profiles and the rotation vector model of the
position angle \citep{mgs11}.

\section{Radio magnetars with GPS}

\subsection{GPS and turn-up in J1550$-$5418}

The flux density measurements of PSR J1550$-$5418 have been obtained using ATCA and the Parkes radio telescope (Australia)
\citep{cam07, cam08, bates11} and the spectrum of this magnetar is presented in Fig. 1. At each frequency for which multiple
observations exist, the variability appears to be of order $30\%$ about the mean, but can reach $50\%$ \citep{cam08}. We
present average values of flux measurements from the available published data. This spectrum looks like GPS and, in addition,
at high frequencies it might be showing a turn-up, as it is observed in the spectra of normal pulsars \citep{wieleb93,
kram96}. Let us note once again that \citet{cam08} suggested that the spectrum of this pulsar could be a turnover with the maximum at 6 GHz.

\subsection{PSR J1622$-$4950}
For the frequency ranges 1.4 $-$ 9 GHz and 17 $-$ 24 GHz we have used data obtained by \citet{lev10} and
\citet{keith11}, respectively. The values at 1.4 and 3.1 GHz are statistical averages, whereas all the values at higher
frequencies are obtained from single observations. \cite{and12} presented the fluxes at 5 and 8 GHz from ATCA
observations in November and December of 2008. These measurements are greater by a factor $\sim 2.5$ than those
presented in Fig. 1. The flux measurements at 5 and 9 GHz presented in Fig. 1 were obtained between 2009 and 2010
\citep{lev10}. This discrepancy is caused by the strong radio variability which may be a direct result of the X-ray
outburst occurred in 2007 \citep{and12}. Since the discovery of this radiomagnetar in April 2009 the same phenomenon
has been observed at 1.4 GHz by the Parkes telescope.

The collected data of the integrated flux density gathered since the discovery is plotted in the bottom panel of Fig.~1
in \cite{lev12}. Simply fitting a line to the data points results in a slight slope, giving a decline of the average
flux density of $\sim 2$ mJy for the 700 days of observing \citep[see Fig. 1 in][]{lev12} suggesting, therefore, an
intrinsic long-term decay of the flux density. That is why we present the spectrum when the pulsar flux achieves
stability.

\begin{table}
\caption{The results of our fits to the radio magnetars spectra. $\nu_{\rm p}$ is the peak frequency estimated using our
fitting porecedure (see Eq.~1). \label{table_fit}}
\begin{center}
\begin{tabular}{ccrrcc} \hline PSR & a & \multicolumn{1}{c}{b} &
\multicolumn{1}{c}{c} & $\chi^2$ & $\nu_{p}$ \\
& & & & & (GHz) \\
\hline
J1550$-$5418 & -1.0$\pm 0.2$ & 1.4$\pm 0.3$ & 0.35$\pm 0.08$ & 1.85 & 5.0 \\
& & & & \\
J1622$-$4950 & -1.8$\pm 0.3$ & 3.3$\pm 0.6$ & -0.35$\pm 0.32$ & 0.91 & 8.3 \\
\hline
\end{tabular}
\end{center}
\end{table}

\subsection{The fitting procedure and data analysis}
To analyze spectral shapes
we use the fitting method previously used by \citet{kuzm01} for pulsar spectra with a turnover at low frequencies. The same
method was used by \citet{kijak11b} in the case of GPS pulsars and B1259$-$63. Based on the data of flux measurements with errors for these two radiomagnetars we have characterized the spectral shapes as
\begin{equation}
\label{flux_func}
F(\nu)=10^{~\left(ax^2+bx+c\right)},\ \ \ {\rm here}\ x \equiv \log_{10}{\nu}.
\end{equation}
Here $\nu$ is the observed frequency and $a$, $b$ and $c$ are the fitting parameters. In order to fit the data we used
an implementation of the nonlinear least-squares Marquardt-Levenberg algorithm. Table~\ref{table_fit} summarizes the
results of fitting of function (1) to the radio-magnetar spectra with the peak frequencies. Table~\ref{table_fit} also quotes
the obtained values of normalized $\chi^2$ which show that the fit to the J1622$-$4950 spectrum
is better than the fit to the J1550$-$5418 spectrum. Thus, both radiomagnetars clearly show GPS feature with the peak frequencies at 5.0 and 8.3 GHz, respectively (see table~\ref{table_fit}).

\begin{table}
\caption{Pulsars with GPS, radiomagnetars and the case of PSR B1259$-$63. $\nu_{\rm p}$ is estimated using different
spectral fit methods \citep{kijak11a,kijak11b}, i.e. the values of the peak frequency quoted in brackets are from
two-power-law
model fit, other values are from the non-linear function fit (see. Eq.~\ref{flux_func}). For B1259$-$63 we quote the
value of $\nu_{\rm p}$ for three different ranges of orbital phases (expressed as the days prior/past the periastron passage). \label{table_peak}}
\begin{center}
\begin{tabular}{lcccc}
\hline
PSR & DM & Age & $\dot E$ & $\nu_{\rm p}$ \\
& $\left ({\rm pc~cm}^{-3}\right)$ & (kYr) & (erg/s) & (GHz) \\
\hline
B1054$-$62 & 320 & 1870 & 1.9e+33 & 0.7(1.0) \\
J1809$-$1917 & 197 & 51 & 1.8e+36 & 2.3(1.7) \\
B1822$-$14 & 357 & 195 & 4.1e+34 & 1.4(1.4) \\
B1823$-$13 & 231 & 21 & 2.8e+36 & 1.7(1.6) \\
B1828$-$11 & 161 & 107 & 3.6e+34 & 0.7(1.2) \\
\hline
\multicolumn{5}{c}{Binary system with Be star LS 2883} \\
B1259$-$63 & 146 & 332 & 8.2e+35 & --- \\
\multicolumn{2}{c}{ $-$60:$-$40 days }&&& 2.5 \\
\multicolumn{2}{c}{16:20 days} &&& 3.4 \\
\multicolumn{2}{c}{63:94 days} &&& 1.5 \\
\hline
\multicolumn{5}{c}{Radiomagnetars} \\
J1550$-$5418 & 830 & 1.4 & 1.0e+35 & 5.0(6.5)\\
J1622$-$4950 & 820 & 4.0 & 8.5e+33 & 8.3(9.0) \\
\hline
\end{tabular}
\end{center}
\end{table}

\section{Discussion}

The pulsed radio emission from a magnetar was first detected in 2006 by the Parkes telescope while observing the position of
J1810$-$197 \citep{cam06}. Its radio spectrum in the range of frequencies between 1.4$-$32 GHz is described by a single, flat
power law with the spectral index $\alpha=0.0\pm 0.5$ \citep{laza08}. Only two more radio-emitting magnetars
(J1550$-$5418 and J1622$-$4950) that are the subjects of our interest have been found up to now. Most probably both of them
are associated with supernova remnants. Radio observations reveal that both magnetars are located in the center of
radio shells emitting a faint diffuse radio emission \citep{and12, gelf07}. This environment consisting of hot,
ionized gas should influence the magnetar radio emission. Therefore we can expect that the magnetar radio-spectra can be subject to the same kind of changes as the spectra of the GPS pulsars. It is also possible that the spectrum of
J1622$-$4950 is influenced by a nearby HII region \citep{and12}. The typical GPS pulsar PSR B1054-62 lies behind or within a dense HII region \citep{kori95} and its spectrum shows somewhat less power at low frequencies \citep{kijak11a}. The similar effect can be noticed while comparing the spectra of J1550$-$5418 and J1622$-$4950 and the difference between these spectra can be caused by HII region \citep[see Figure 2 in][]{and12}.

Table~\ref{table_peak} summarizes the results of peak frequency estimations for the following sources: the isolated
GPS pulsars, the PSR~B1259$-$63 for three orbital phase ranges (for a full list see \citealt{kijak11b}) and two
radio-magnetars. As it can be seen the peak frequency for the regular radio pulsars ranges from 0.7 to 2.3~GHz, while in the case of B1259$-$63 it reaches 3.5~GHz close to the periastron passage. Let us note, that the spectrum evolves to a regular power-law pulsar spectrum for those orbital phases where the pulsar is far from its companion star. 
The values of $\nu_{\rm p}$ which we have obtained for the radio magnetars are much higher, 5 and 8~GHz. But the question arises, why do not we observe regular radio pulsars with the turnover at such high frequencies? This question seems natural since we suggest that
for both regular pulsars and radio magnetars the GPS appearance is due to the environmental conditions around the
neutron stars.

There are two possible explanations: either the magnetar environments are even ``more extreme'' than those for the GPS
pulsars or, more plausibly, pulsars with turnovers at such high frequencies have not been discovered yet. Most of the
regular pulsar search surveys were conducted at low frequencies usually 1.4~GHz or below. A pulsar with a peak frequency of a
few gigahertz would be extremely weak at these frequencies, hence probably undetectable. As for the radio emissions of those
two magnetars they were found only after the sources had been identified by other means. Obviously, the detection of pulsed
radio emission is much easier when the period of rotation has been already known than its detection in {\it a blind survey}.
Additionally, the pulse broadens due to interstellar scattering and this effect is severe for high DM pulsars
\citep[see][]{kijak11a}.

Only a few pulsar surveys have been conducted at higher frequencies, where it is easier to find ``the
missing link'' -- a pulsar with a turnover at the frequency of a few GHz. Two of the most recent surveys, 
conducted at 3 \citep{keith08} and 6~GHz \citep{bates11}, yielded a few promising discoveries of pulsars that can be potential
candidates for GPS sources. Of course, these pulsars, especially their spectra, still require further investigation,
what we intend to do in our future observing projects for GPS pulsar candidates.

At present it is difficult to construct a detailed theory of the turnover spectra as we lack observational data especially 
that of the GPS sources. However, from the available observational data we can still draw a general conclusion
about the main factors that influence the observed spectra. It is generally known that absorption at low
frequencies is caused by means of thermal absorbtion in the hot stellar wind \citep{sieb73,slee86}. Preliminary estimations
show that free-free absorption in to PWNs, HII regions can be responsible for this effect. The optical depth
$\tau_{\nu}$ of the free-free absorption can be expressed as \citep{Ryb79}
\begin{equation}
\tau_{\nu}=0.4\times
T_{3}^{-3/2}\nu^{-2}_{\rm GHz}\int n_{e}^{2}dl_{\rm AU}~,
\end{equation}
where $n_e$ is the electron density in cm$^{-3}$, $T_3$ is
the temperature in $10^3$ K, $\nu_{\rm GHz}$ is the frequency in GHz and $l_{\rm AU}$ is the distance in AU or
according \citet{rolfs96} and \citet{wilson09}
\begin{equation}
\tau_{\nu}=8.235\times 10^{-2} \left(\frac{T_{e}}{\rm K}\right)^{-1.35} \left(\frac{\nu}{\rm GHz}\right)^{-2.1}
\left(\frac{\rm EM}{\rm pc~cm^{-6}}\right)~a(\nu,T)~,
\end{equation}
where the correction $a(\nu,T)$ is usually $\sim 1$ and the {\it emission measure} EM is given by
\begin{equation}
\left(\frac{\rm EM}{\rm pc~cm^{-6}}\right)=\int\limits^{l/pc}_{0} \left(\frac{N_{e}}{\rm cm^{-3}}\right)^{2}
d\left(\frac{l}{\rm pc}\right).
\end{equation}
When the frequency is given in units of $\nu_0$ , where $\nu_0$ is that frequency at which the optical depth is unity,
we have, from (3):
\begin{equation}
\frac{\nu_0}{\rm GHz}=0.3045 \left(\frac{T_{e}}{\rm K}\right)^{-0.643} \left(\frac{a(\nu,T)\rm EM}{\rm
pc~cm^{-3}}\right)^{0.476}.
\end{equation}
{Thus, we can conclude that the GPS feature can be caused by external factors, in the same way as it occurs in the
environment of normal GPS pulsars. The GPS pulsars apparently are surrounded by some kind of environment that can affect the
spectra of those pulsars in the same way as the stellar wind affects the B1259$-$63 spectrum. For example, PSR B1823$-$13 is
surrounded by a compact pulsar wind nebula \cite{pav11} and PSR B1054-62 lies behind or within a dense HII region
\citep{kori95}.

\section{Conclusions}

As it follows from the discussion above the average spectra of two radiomagnetars, J1550$-$5418 and J1622$-$4950, possess characteristic features of GPS. 
In the Introduction we have already discussed that the coherent curvature radiation should be the mechanism of radio
emission for both magnetars and normal pulsars. Despite a substantial difference between the observed radio features of
magnetars and normal pulsars, both of them emit highly linearly polarized individual pulses. We believe that this essential
feature unambiguously points to the similar emission mechanism. Therefore, we conclude that the GPS feature of the magnetars
is caused by their environment. Thus, the radiomagnetars PSR J1550$-$5418 and J1622$-$4950 can join the class of the GPS
pulsars.

\acknowledgments We are grateful to the anonymous referee for useful comments. 
This paper was supported by the grant DEC-2012/05/B/ST9/03924 of the Polish National Science Centre. 
We thank M.Margishvili for the language editing of the manuscript.

\end{document}